\documentclass[12pt,cites]{article}
\usepackage{latexsym,graphicx,amssymb,amsmath}
\usepackage{cite}
\author{A. V. B.  Cruz$^a$,  A. K. Mishra$^a$ 
and W. Schmickler$^b$\\
$^a$ Institute of Mathematical Sciences\\CIT Campus, Chennai, 600113 India\\
$^b$ Institute of Theoretical Chemistry, Ulm University\\
D89069 Ulm, Germany}
\title{Electron tunneling between two electrodes mediated by  a molecular wire containing a redox center}
\begin{document}
\maketitle

\begin{abstract}
We derive an explicit expression for the quantum conductivity of a molecular wire containing a redox center, which is embedded 
in an electrochemical environment. The redox center interacts with the solvent, and the average over the solvent configurations is 
performed numerically. Explicit calculations have been performed for a chain of three atoms. When the redox center interacts strongly with neighboring electronic levels, the current-potential curves show interesting features like rectification, current plateaus and negative differential resistance. Electronic spectroscopy of intermediate states can be performed at constant small bias by varying the electrochemical potential of the wire.
\end{abstract}

\section{Introduction}

Understanding of electron transport through a single molecules received an increased interest due to the speculation of employing molecular units as fundamental elements of computer circuits \cite{Jort_ed,Aviram_ed}. Additionally, electron transfer in molecular wires at nanoscale level received further attention, both at the level of formalism as well as ab-initio calculations, due to its possible relevence in understanding and application for a class of diverse problems like sensors, photonics, solar energy conversion \cite{Adams01}. Controlled charge movement in a suitably designed molecule can be used as the basis for storing and processing of information. Such Quantum-Dot Cellular Automata architecture has been experimentally realised in a series of experiments for a variety of applicable components like memory cells, logic gates and clocked memory cells\cite{lent01,Amlani01,Kum01}. A practical implementation of QCA architecture consists of a single redox center with an organic or inorganic bridging group \cite{Lieberman01}. The efficiency of solar energy conversion process depends  not only on efficient photon capture but also on charge seperation and transport through very large distances. Since the charges are created by sunlight on the surface of an assembly of molecules or semiconductors, it is resonable to expect molecular wires to act as relevant acceptors of the charges. Also the weak solar fluxes imply a very low  current and a need for fast charge transport. Certain classes of polymers and oligomers have been proposed as ideal candidates for satisfying the above criterion for increasing the yield in solar energy conversion \cite{cygan01,Decher01}.  

 The above mentioned are some of the reasons for the recent increased surge in interest for understanding charge transport along molecular wires. Typical theoretical work in the field  involves obtaining generic expressions for  the  conductance, current-voltage  profiles, rate constants, transition probabilites etc. Formal works on transport properties along molecular wires were carried out on Donor-Bridge-Acceptor (DBA complexes) systems, wherein electrons are transferred  between donor and acceptor connected by a molecular bridge \cite{Newton01,Davis01,Marcus01,McConnell01}. Further, electron transfer between reservoirs connected by a molecular bridge has been studied by Ratner and co-workers \cite{Mujica01,Mujica02,Mujica03}. The above works resort to time-dependent quantum mechanics for obtaining expressions relevant to electron transfer. Expressions for the conductance between two reservoirs connected by monoatomic sites were well known in mesoscopic physics. The formal expression was first derived by Caroli and co-workers \cite{caroli} and  was later expanded to a broader class of problems by Wingreen and co-workers \cite{wingreen1,meir1,meir2}. Recently, the same expressions were re-derived by various authors \cite{Hu01,Camalet01,Camalet02,Segal01,sen01} by formulating Quantum Langevin Equations (QLE).   Initially, both the molecular wire and mesoscopic conduction were modelled using tight-binding Hamiltonians, and since at a Hamiltonian level these problems seem identical, it is expected that the expression obtained for one should be applicable for the other.  

  Several  authors have pursued other computational methods such as density functional theory, first principle ab-initio calculations and package simulation of Non-Equilibrium Green's Function in studying conduction through molecular wires \cite{Brand01,Taylor01,Taylor02,DiVentra01,Datta02}. Most of these authors differ in their treatment of the electrodes and the interaction of the metal-molecule coupling. Most of the earlier works were oriented towards a better approximation for modelling the interaction and self-energies at the molecule-metal junctions,  while little work has been done in incorporating the effect of additional interactions the electron might experience in the molecular wires. Though numerous works have been done of the subject of electron-phonon coupling with relevance to quantum dots \cite{wingreen1,Lundin01,Zhu01,Alexandrov01,Alexandrov02,Flensberg01,Mozyrsky01}, including treatments for classical, quantum, equilibriated and out of equilibrium systems, exact treatment of such a process in specific for molecular wire has not received much attention. Molecular wires differ from quantum dots in that the observed conductance behaviour of quantum dot is dominated by Coloumb blockade.

The present work  focuses on a special electrochemical case: a molecular wire containing a redox-center connecting two  electrodes. The electrochemical case is of special interest since two potentials can be varied independently: the bias between the two electrodes, and the potential of one of the electrodes with respect to the solution. The latter acts like a gate voltage that controls the current in the wire. In addition, the redox center interacts with the solvent, whose fluctuation will affect the current. 
The specialties of the electrochemical situation were first elucidated by theorists \cite{me1,cindra,ulstrup1,me2}. 
Starting with the pioneering paper of Tao \cite{Tao01}, there have been a fair number of experimental studies of electrochemical systems 
\cite{Han01,Holmin01,Tran01,Xiao01,He01,Albrecht01,Chi01,Li01,Alessandrini01}
which in turn have generated more theoretical work (see e.g. \cite{galperin,medved} and references therein). 

Most of the theoretical work on electrochemical systems has been restricted to special systems with one or two intervening redox centers. In this study, we will consider a wire of arbitrary lengths containing one redox center interacting with the solvent. Using a tight-binding Hamiltonian and Green's function techniques we will derive an expression for the current which is exact for the case where the interaction to the two electrodes can be treated in the wide-band approximation. These calculations will be illustrated by model calculations for particularly interesting cases: steps and negative differential resistance, and spectroscopy  of intermediate electronic states.

\section{Model Hamiltonian, Green's functions and current density}
The model system that we consider consists of two metal electrodes, labeled R and L, connected by a chain of $2n +1$ atoms -- an odd number is chosen for convenience only. The atom in the center 
is redox-active and interacts with the solvent; thus, we identify the index $n+1$ with the index $r$ of the redox species. We use a tight-binding model, in which each atom contains one orbital and interacts only with its nearest neighbor. The corresponding Hamiltonian can be written in the form:  
  
\begin{align}\label{hamil}
\mathbf{H} & = \sum_k \sum_{i=L,R} \epsilon_{k,i} n_{k,i} + \epsilon_{r} n_{r} + \sum_{\substack{i =1 \\i \neq n+1}}^{2n+1} \epsilon_{i} n_{i} \nonumber  \\
 & \qquad  + \sum_{i=1}^{n-1} \lbrace \upsilon_{i}c_{i}^{\dagger}c_{i+1} + h.c \rbrace  + \sum_{i=n+2}^{2n} \lbrace \upsilon_{i} c_{i}^{\dagger}c_{i+1} + h.c \rbrace \nonumber   \\
 &  \qquad  +\sum_{i=n,n+1(r)} \lbrace \bar{\upsilon_{i}}c_{i}^{\dagger}c_{i+1} + h.c \rbrace  + \sum_{k} \lbrace v_{k,1}c_{k,L}^{\dagger}c_{1} + v_{k,2n+1}c_{k,R}^{\dagger}c_{2n+1} + h.c \rbrace \nonumber   \\
 &  \qquad  + \frac{1}{2}\sum_{\nu} \hbar \omega_{\nu} q_{\nu}^{2} + \sum_{\nu} \hbar \omega_\nu g_{\nu} q_{\nu} n_{r}
\end{align}

In this Hamiltonian, $n$ always denotes an occupation number,  $c^{\dagger}$ a creation $c$ an annihilation operator, $\epsilon$ an energy, and $v$ a coupling constant. The first line contains the diagonal elements, the indices $(k,L)$ and  $(k,R)$ labeling the electronic states on the two electrodes. 
The second and third lines give hopping elements between adjacent sites, and the last line the potential energy of the solvent, with coordinates $q_\nu$ and frequencies $\omega_\nu$, and its interaction 
with the redox center $r$; the $g_\nu$ are the corresponding coupling constants. Equation (\ref{hamil}) is a natural generalization of the Hamiltonian for redox-mediated tunneling via one center \cite{me2}.

 The matrix form of the fermionic part of the above Hamiltonian $H_{F}$ has the generic form:

\begin{displaymath}
\mathbf{H_{F}} = \left( \begin{array}{ccc}
\epsilon_{k,L} & v_{k} & 0 \\
v_{k} & \mathbf{H}_{chain} & v_{k} \\
0 & v_{k}  & \epsilon_{k,R} \\
\end{array} \right)
\end{displaymath}

\begin{displaymath}
\mathbf{H}_{chain} = \left( \begin{array}{ccccccc}
\epsilon_{1} & \upsilon &&&&& \\
\upsilon  & \epsilon_{2} & \ddots &&&& \\
&\ddots&& \bar{\upsilon} & &&\\
  &&  \bar{\upsilon}  & \epsilon_{r} + \lambda_{\nu}q_{\nu} & \bar{\upsilon} & & \\
&&& \bar{\upsilon}  &&\ddots&\\
  &&&&\ddots&& \upsilon \\
   &&&&& \upsilon & \epsilon_{2n+1} \\
\end{array} \right)_{2n+1 \times 2n+1}
\end{displaymath}

Working within the tight-binding model, it is understood that $\mathbf{H}_{F}$ has  non-zero entries only in  diagonal and sub-diagonal elements. The general scheme of approach is to calculate the quantum conductance and then obtain the current by integrating the conductance between appropriate limits. The formula employed for obtaining the quantum conductance, or tunneling rate, is the same as the one used by Datta {\it et al.} \cite{Datta1}. This form of the formula was first derived by Caroli {\it et al.} \cite{caroli} and was subsequently derived in a much wider context by Wingreen and co-workers \cite{wingreen1,meir1,meir2}.

\begin{align}
g = {\mathrm Tr} \lbrack {\mathrm G}^{r} \varGamma_{L} {\mathrm G}^{a} \varGamma_{R} \rbrack 
\end{align}

As before, the subscripts $L$ and $R$ refer to the left and right reservoir. $\varGamma$ denotes the imaginary part of self-energy (for ease of following the notations, $\varGamma = v_{k} (Im G^{0} _{kk} ) v^{*} _{k} $).  The quantity of interest  is  $ \mid \langle 1 \mid G \mid 2n+1 \rangle \mid ^{2}$, where $G$ is the Green's function obtained from the above Hamiltonian. This can be obtained by separating the Hamiltonian into two parts: $\mathbf{H} = \mathbf{H}_0 + V$ and considering a Dyson equation , $ G = G^{0} + G^{0}VG $, where the simplification of the problem results from the choice of $V$. Letting $ V = \sum_{i=n,n+1} \bar{\upsilon}_{i}c_{i}^{\dagger}c_{i+1} + h.c $, the Hamiltonian $\mathbf{H}_{F}^{0}$ contains 3 block matrices. Physically this amounts to cutting the $2n+1$ atom chain at 2 places on either side of the redox couple.  The closed form for the element $\langle 1 \mid G \mid 2n+1 \rangle$ is obtained as shown:

\begin{align}
\langle 1 \mid G \mid 2n+1 \rangle & =  \langle 1 \mid G^{0} \mid 2n+1 \rangle + \sum_{i,j} \langle 1 \mid G^{0} \mid i \rangle\langle i \mid V \mid j \rangle \langle j \mid G \mid 2n+1 \rangle  \\ \nonumber
& =     \langle 1 \mid G^{0} \mid n \rangle\langle n \mid V \mid n+1 \rangle \langle n+1 \mid G \mid 2n+1 \rangle 
\end{align}

 \begin{align}
\langle n+1 \mid G \mid 2n+1 \rangle & =  \langle n+1 \mid G^{0} \mid n+1 \rangle \langle n+1 \mid V \mid n \rangle \langle n \mid G \mid 2n+1 \rangle   \\ \nonumber
&  + \langle n+1 \mid G^{0} \mid n+1 \rangle  \langle n+1 \mid V \mid n+2 \rangle \langle n+2 \mid G \mid 2n+1 \rangle 
\end{align} 

 \begin{align} 
\langle n \mid G \mid 2n+1 \rangle =  \langle n \mid G^{0} \mid n  \rangle \langle n \mid V \mid n+1 \rangle \langle n+1 \mid G \mid 2n+1 \rangle 
\end{align}

 \begin{align} 
\langle n+2 \mid G \mid 2n+1 \rangle&  =  \langle n+2 \mid G^{0} \mid 2n+1 \rangle \\ \nonumber +  \ & \langle n+2 \mid G^{0} \mid n+2 \rangle \langle n+2 \mid V \mid n+1 \rangle \langle n+1 \mid G \mid 2n+1 \rangle 
 \end{align} 

 From the above 4 equations , $G_{1,2n+1} = \langle 1 \mid G \mid 2n+1 \rangle$ can be solved:
\begin{equation}
\label{gbe}
 G_{1,2n+1} = \frac{G^{0}_{1,n} \bar{\upsilon}_{n,n+1}G^{0}_{n+1,n+1}\bar{\upsilon}_{n+1,n+2}G^{0}_{n+2,2n+1}}{1 - G^{0}_{n+1,n+1}\lbrack \bar{\upsilon}_{n+1,n}G^{0}_{n.n}\bar{\upsilon}_{n,n+1} + \bar{\upsilon}_{n+1,n+2}G^{0}_{n+2,n+2}\bar{\upsilon}_{n+2,n+1} \rbrack} 
\end{equation}

Now we require the terms $ G^{0}_{1.n},G^{0}_{n+1,n+1},G^{0}_{n+1,n+2},G^{0}_{n+2,2n+1},G^{0}_{n+2,n+2}$. These can be found by using the above reduced Green's function technique, in addition to exploiting the recursive relation for the determinant of a matrix consisting only of diagonal and sub-diagonal entries. Similar calculational methods were employed by Evenson and Karplus \cite{evenson}.

For simplicity, we assume that the couplings to the two metals at the ends are the same, and use the wide-band approximation, in which $\Delta = \pi \sum_k v_k^2 \delta(\epsilon - \epsilon_k)$ is taken as constant.

 If $d_{n}$ represents the determinant of a $n \times n$  matrix with diagonal entries set to $\epsilon - \epsilon_{i}$ and subdiagonal entries set to some $\upsilon$, then it is possible to express:

\begin{align} G^{0}_{1,n} = \frac{(-\upsilon)^{n-1}}{ d_{n} - d_{n} \lbrack \frac{v_{k}^{2}}{\epsilon - \epsilon_{k}} \rbrack } \end{align} 

\begin{align} G^{0}_{n,n} = \frac{d_{n-1} -  d_{n-2} \lbrack \frac{v_{k}^{2}}{\epsilon - \epsilon_{k}}\rbrack}{d_{n} - d_{n-1} \lbrack \frac{v_{k}^{2}}{\epsilon - \epsilon_{k}} \rbrack }  \end{align} 

\begin{align} G^{0}_{n+1,n+1} = \frac{1}{\epsilon - \epsilon_{r}}  \end{align} 

 \begin{align} G^{0}_{n+2,2n+1} \cong G^{0}_{1,n}  \end{align} 

\begin{align} G^{0}_{n+2,n+2} \cong G^{0}_{n,n}  \end{align}

Invoking the wide band approximation the above expressions reduce to the following form:

\begin{align} G^{0}_{1,n} = \frac{(-\upsilon)^{n-1}}{ d_{n} + i d_{n} \Delta} \end{align} 

\begin{align} G^{0}_{n,n} = \frac{d_{n-1} + i d_{n-2} \Delta }{d_{n} + i d_{n-1} \Delta } \end{align} 

Thus, the problem has been reduced to calculating the determinants $d_n$, which is given in the appendix.

 The conductance can be obtained from Caroli's formula and integrated to obtained the net current. The net current thus obtained has a $q_{\nu}$ dependency which has to be eliminated by performing a thermal averaging.   The final result obtained after thermal averaging gives the net total current.
 As noted in the introduction, in electrochemical systems there are two potential differences to consider: the bias $V$ between the two electrodes, and the electrode potential, which shifts the levels in the solution.
 We use the convention that the potential of the right electrode $R$ is kept constant, and set its Fermi level to zero. 
The levels on the wire shift with the electrode potential; this assumes that the conductivity of the solution is higher than that of the wire. Other scenarios can be calculated by the same formalism. 
With this convention, we write the total current in the form:
\begin{equation}
\label{iofq}
 I(q) = \int {\mathrm Tr} \lbrack {\mathrm G}^{r} \varGamma_{L} {\mathrm G}^{a} \varGamma_{R} \rbrack
   \{ f(\epsilon + e_0V) - f(\epsilon) \}  d\epsilon 
\end{equation}
where  $f(\epsilon)$ denotes the Fermi-Dirac distribution. As has been pointed out several times
(see e.g. \cite{met1}), in the case of a classical solvent it is sufficient to consider a single effective solvent coordinate $q$. Effectively, this means that in the Hamiltonian, we make the following substitutions:
\begin{equation}
\label{subs}
\frac{1}{2} \sum_\nu \hbar \omega_\nu a_{\nu}^2\ \to \ \lambda q^2, \qquad  \sum_\nu \hbar \omega_\nu g_\nu q_\nu \ \to \ -2 \lambda q 
\end{equation}

The average over the solvent configurations can then be written as:

\begin{equation}
\label{ }
 I = \frac{1}{Z} \int dq e^{-\beta E(q) } I(q) \qquad Z = \int dq e^{-\beta E(q) }
\end{equation}
where the energy, as a function of the solvent coordinate $q$, is:

\begin{equation}
\label{eofq}
E(q)  = \lambda q^{2} + \sum_{i} \int \epsilon \langle c_{i} ^{\dagger} c_{i} \rangle 
\end{equation}

 Thus $E(q)$ is obtained by performing a partial trace over the fermionic part of the total Hamiltonian. The quantity $\langle c_{i} ^{\dagger} c_{i} \rangle$, as viewed by Wingreen et al. \cite{wingreen1}, is the lesser component of the Keldysh Green's function, $\mathrm{G}^{<} _{ii}$. If $\varGamma_{L}$ 
 and $\varGamma_{R}$ are the imaginary parts of the self-energy arising from the interaction with the left and the right reservoirs (which in the view of wide-band approximations is $\Delta$), then for the present case 
\begin{equation}
\label{g1}
 \mathrm{G}^{<} _{ii} = i f (\epsilon + e_0V) \lbrack {\mathrm G}^{r} \varGamma_{L} {\mathrm G}^{a} \rbrack _{ii} + i f (\epsilon) \lbrack {\mathrm G}^{r} \varGamma_{R} {\mathrm G}^{a} \rbrack _{ii}  
\end{equation}

At this point, a few comments on the appearance for ${\mathrm G}^{<}$ are needed. It is well known that in equilibrium the lesser Green's function takes the form of a product of spectral function times the occupation function. (${\mathrm G}^{<} = i a(\epsilon)f(\epsilon) $). That is in case of zero bias $V$ when both the reservoirs have the same potential, $f(\epsilon+e_0V) = f(\epsilon) $, then 
\begin{equation}
\label{g2}
{\mathrm G}^{<} = i f(\epsilon) \lbrack {\mathrm G}^{r} ( \varGamma_{L} + \varGamma_{R} )  {\mathrm G}^{a} \rbrack = i f(\epsilon) 2\Delta \lbrack {\mathrm G}^{r} {\mathrm G}^{a} \rbrack  
\end{equation}

   Now ${\mathrm G}^{r} (\epsilon) - {\mathrm G}^{a} (\epsilon)  = a(\epsilon) $ where $a(\epsilon)$ is the spectral function, and the imaginary part of self-energy can be written as $(1/{\mathrm G}^{a} - 1/{\mathrm G}^{r})$. In our notation, $2\Delta = \Delta_{L} + \Delta_{R} = (1/{\mathrm G}^{a} - 1/{\mathrm G}^{r})$. Also the diagonal part of ${\mathrm G}^{r} (\epsilon) - {\mathrm G}^{a} (\epsilon)$ is proportional to the density of states, $\rho(\epsilon)$. In equilibrium, we recover  the result
\begin{equation}
\label{g3}
\mathrm {G}^{<} = i f(\epsilon) a(\epsilon) = f(\epsilon) ({\mathrm G}^{r} - {\mathrm G}^{a}). 
\end{equation}

  Substituting the above result in the expression (\ref{eofq}) for $E(q)$, it is seen that at equilibrium $E(q) = \lambda q^{2} +  \int \epsilon f(\epsilon)  \mathrm {Im Tr} G(\epsilon)d \epsilon$, wherein the second term in the  energy expression is   widely employed  in a variety of contexts in physics. Thus the expression for $E(q)$ for the non-equilibrium case can be written compactly as

\begin{equation}
\label{g4}
E(q) = \lambda q^{2} + \int \epsilon f(\epsilon + e_0 V) \mathrm{Tr} \lbrack {\mathrm G}^{r} \varGamma_{L} {\mathrm G}^{a} \rbrack d\epsilon  + \int \epsilon f(\epsilon) \mathrm{Tr} \lbrack {\mathrm G}^{r} \varGamma_{R} {\mathrm G}^{a} \rbrack d\epsilon  
\end{equation}

The trace in the above equation runs over the $(2n+1)$ sites numbered by the index $i$. Now for better exposition of the computation involved in calculating the $E(q)$, we consider a generic term which has the form  shown below:
\begin{equation}
\label{g5}
E_{i} (q) = \Delta \int_{-\infty} ^{0}  \epsilon {\mathrm G}^{r} _{i,1} {\mathrm G}^{a} _{1,i} d \epsilon + \Delta \int_{-\infty} ^{- eV}  \epsilon {\mathrm G}^{r} _{i,2n+1} {\mathrm G}^{a} _{2n+1 ,i}  d \epsilon
\end{equation}
where we have replaced the Fermi-Dirac distribution by step function, and taken the Fermi level as zero.

  We consider the case for the three possible locations of i, ($i \leq n, i \geq n+2, i= n+1$). As before the general idea behind the approach to get the matrix elements of G is to resort to a Dyson expansion. The closed form equation so obtained has to be solved to get the relevant terms. The choice of V is same as used before.

\begin{equation}
\label{g6}
{\mathrm G}_{1,i} = {\mathrm G}^{0} _{1,i} + {\mathrm G}^{0} _{1,n} \bar{\upsilon}_{n,n+1} {\mathrm G}_{n+1,i} 
\end{equation}
\begin{equation}
\label{g7}
{\mathrm G}_{n+1,i} = {\mathrm G}^{0} _{n+1,i} + {\mathrm G}^{0} _{n+1,n+1} \bar{\upsilon}_{n+1,n+2} {\mathrm G} _{n+2,i} + {\mathrm G}^{0} _{n+1,n+1} \bar{\upsilon}_{n+1,n} {\mathrm G}_{n,i} 
\end{equation}
\begin{equation}
\label{g8}
 {\mathrm G}_{n+2,i} = {\mathrm G}^{0} _{n+2,i} + {\mathrm G}^{0} _{n+2,n+2} \bar{\upsilon}_{n+2,n+1} G_{n+1,i} 
\end{equation}
\begin{equation}
\label{g9}
{\mathrm G}_{n,i} = {\mathrm G}^{0} _{n,i} + {\mathrm G}^{0} _{n,n} \bar{\upsilon}_{n,n+1} {\mathrm G}_{n+1,i} 
\end{equation}

\noindent {\bf Case I: $i \leq n$ }
\begin{equation}
\label{g10}
{\mathrm G}_{1,i} = {\mathrm G}^{0} _{1,i} + \frac{{\mathrm G}^{0} _{1,n} \bar{\upsilon}_{n,n+1}{\mathrm G}^{0}_{n+1,n+1} \bar{\upsilon}_{n+1,n}{\mathrm G}^{0}_{n,i}}{1 - \lbrack {\mathrm G}^{0} _{n+1,n+1} \bar{\upsilon}_{n+1,n+2}{\mathrm G}^{0} _{n+2,n+2} \bar{\upsilon}_{n+2,n+1} + {\mathrm G}^{0} _{n+1,n+1} \bar{\upsilon}_{n+1,n} {\mathrm G}^{0} _{n,n} \bar{\upsilon}_{n,n+1} \rbrack} 
\end{equation}
\begin{equation}
\label{g11}
{\mathrm G}^{0} _{1,i}  = \frac{ (-\upsilon)^{i-1} d_{n-i}}{d_{n} + i \Delta d_{n-1}}  
\end{equation}
\begin{equation}
\label{g12}
 {\mathrm G}^{0} _{n,i}  = \frac{ (-\upsilon)^{n-i} (d_{i-1} +i d_{i-2} \Delta)}{d_{n} + i \Delta d_{n-1}} 
\end{equation}
\vspace{5mm}

\noindent {\bf Case II: $i \geq n+2$ }
\begin{equation}
\label{g13}
{\mathrm G}_{1,i} = \frac{{\mathrm G}^{0} _{1,n} \bar{\upsilon}_{n,n+1} {\mathrm G}^{0} _{n+1,n+1} \bar{\upsilon}_{n+1,n+2}{\mathrm G}^{0}_{n+2,i}}{1 - \lbrack {\mathrm G}^{0} _{n+1,n+1} \bar{\upsilon}_{n+1,n+2}{\mathrm G}^{0} _{n+2,n+2} \bar{\upsilon}_{n+2,n+1} + {\mathrm G}^{0} _{n+1,n+1} \bar{\upsilon}_{n+1,n} {\mathrm G}^{0} _{n,n} \bar{\upsilon}_{n,n+1} \rbrack} 
\end{equation}

 It is crucial to note at this stage that because of the form of perturbation selected the unperturbed ${\mathrm G}^{0}$ has a symmetric structure with respect to the first and third block matrix and hence ${\mathrm G}^{0} _{n+2,i}$ in the above is same as ${\mathrm G}^{0} _{1,i}$ in case I. 
\vspace{5mm}

\noindent {\bf Case III: $i = n+1$ }
\begin{equation}
\label{g14}
 {\mathrm G}_{1,n+1} = \frac{ {\mathrm G}^{0} _{1,n} \bar{\upsilon}_{n,n+1} {\mathrm G}^{0} _{n+1,n+1}}{1 - \lbrack {\mathrm G}^{0} _{n+1,n+1} \bar{\upsilon}_{n+1,n+2}{\mathrm G}^{0} _{n+2,n+2} \bar{\upsilon}_{n+2,n+1} + {\mathrm G}^{0} _{n+1,n+1} \bar{\upsilon}_{n+1,n} {\mathrm G}^{0} _{n,n} \bar{\upsilon}_{n,n+1} \rbrack}  
\end{equation}

 Even though $\bar{\upsilon} _{n,n+1} = \bar{\upsilon} _{n+1,n+2} = \bar{\upsilon}$, we have maintained the subscript indices for ease of checking the final expressions.  

\section{Results and discussions}
The principle new feature of our work is the dynamic interaction of the redox system with the adjoining species, which fluctuates with the solvent coordinate $q$. The main effects can be demonstrated with a chain of three atoms, and we limit our numerical calculation to this case.

Even though we have restricted our treatment to the symmetric case, in which the coupling $\Delta$ to the two leads and the interatomic couplings $v$ are the same on both sides, the system contains a fair 
amount of parameters. In the following model calculations, we have set $\Delta = 0.3$~eV and the reorganization energy $\lambda = 0.3$~eV unless otherwise mentioned, and for the other parameters we have chosen values appropriate to demonstrate special effects.

The case of a single intervening redox center is well examined. The new feature of the three-atom chain is the interaction between the levels $\epsilon_1$ of the two side atoms and the redox center. Before considering this in detail, it is instructive to investigate the reference case in which this effect is weak; in this limit, we should obtain similar results to the case of one atom.

\begin{figure}[htb]
\begin{center}
\includegraphics[height=7cm]{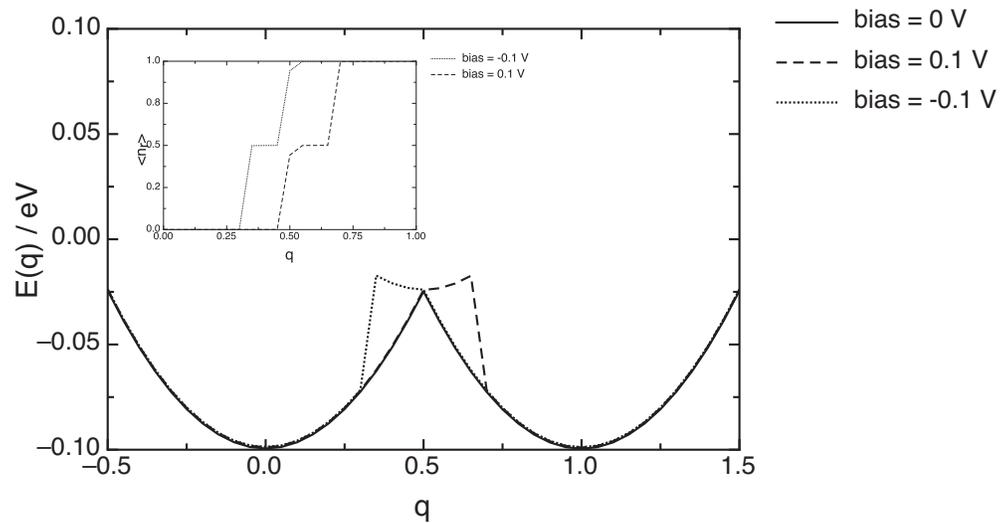}
\caption{Potential energy surfaces in the case of weak coupling and at the equilibrium potential for the redox system; system parameters: $\epsilon_1 = 0.8$~eV, $v=0.01$~eV, $\epsilon_r = \lambda = 0.3$~eV. The insert shows the occupation $\langle n_r \rangle$ of the redox center.}
\label{poteq}
\end{center}
\end{figure}

The interaction of the side atoms with the center is weak, if $\epsilon_1$ lies far from the Fermi level  and the coupling $v$ is small.  In this limit, the redox center is at the equilibrium potential for $\epsilon_r = \lambda$. As expected, in this case the potential energy surfaces $E(q)$ are similar to the one atom case \cite{sashame}. At zero bias, they  have the same form as for a normal, outer sphere redox reaction- see Fig. \ref{poteq}. In the left well, the occupancy $\langle n_r \rangle$ is zero, in the right well unity. Application of a bias produces a region with $\langle n_r \rangle \approx 1/2$, which extends the barrier in the center. For the three atom case, the tunneling rate, as a function of the electronic energy $\epsilon$  and the solvent coordinate $q$, is given by:
\begin{equation}
\label{tunnel}
t(\epsilon,q) = \frac{\Delta^2  \bar{v}^4}{(\epsilon - \epsilon_r + 2 \lambda q)^2 \left[(\epsilon - \epsilon_r + 2 \lambda q)(\epsilon-\epsilon_1 + i \Delta) - 2 \bar{v}^2\right]^2
}
\end{equation}
As long as $\epsilon_1$ lies so high that it plays no role, this rate has a maximum where $
\epsilon - \epsilon_r + 2 \lambda q = 0$. Tunneling occurs only between the two Fermi levels,
in the range $- e_0 V < \epsilon < 0$. Inspection shows, that the maximum of $t(\epsilon, q)$ is obtained in the region where $\langle n_r \rangle \approx 1/2$, which therefore gives the main contribution to the current.

\begin{figure}[hbt]
\begin{center}
\includegraphics[height=7cm]{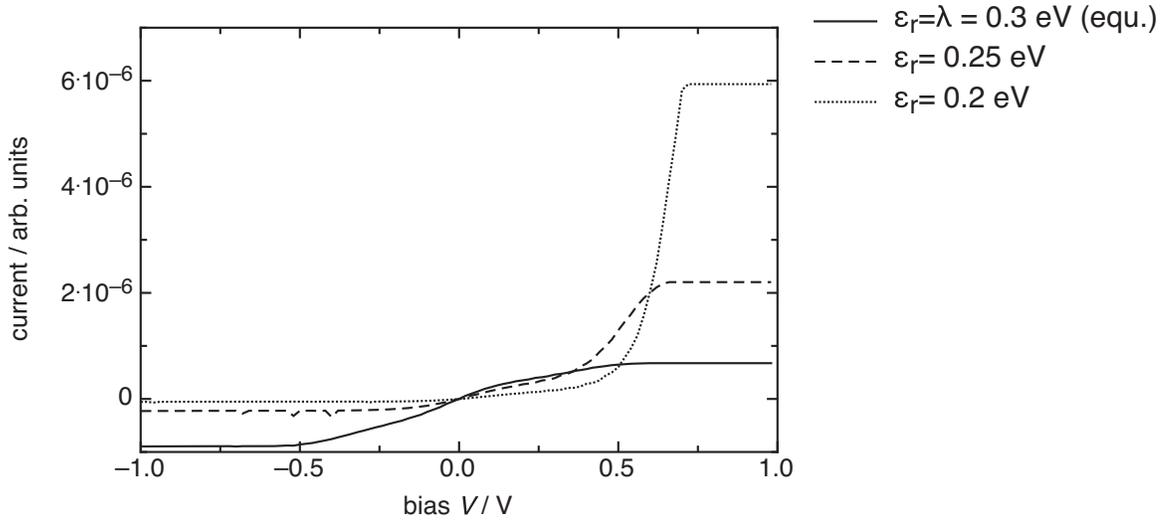}
\caption{Current-potential curves in the case of weak coupling; system parameters: $\epsilon_1 = 1.2$~eV, $v=0.01$~eV, }
\label{iv1}
\end{center}
\end{figure}

Continuing with the case of weak coupling, the current-potential curves are symmetric at the equilibrium
condition $\epsilon_r = \lambda$. Shifting $\epsilon_r$ by application of an overpotential leads to asymmetrical curves with rectifying properties. Figure \ref{iv1} shows the case in which the redox level has been lowered, so that the most favorable energy range now lies below the Fermi level of the right electrode. Therefore the current is higher at positive bias, where this energy range lies between the two Fermi levels.

Really new features occur when the redox level interacts noticeably  with the levels $\epsilon_1$. In general, three interacting atomic levels combine to form three molecular orbitals. Since the redox level changes its energy with the   solvent fluctuations, so do the resulting molecular orbitals. So, for some range of $q$ the redox level will be far from $\epsilon_1$ and the interaction will be  almost negligible, in another range it will lie close in energy, so that one observes the typical splitting of the levels. 

\begin{figure}[htb]
\begin{center}
\includegraphics[height=7cm]{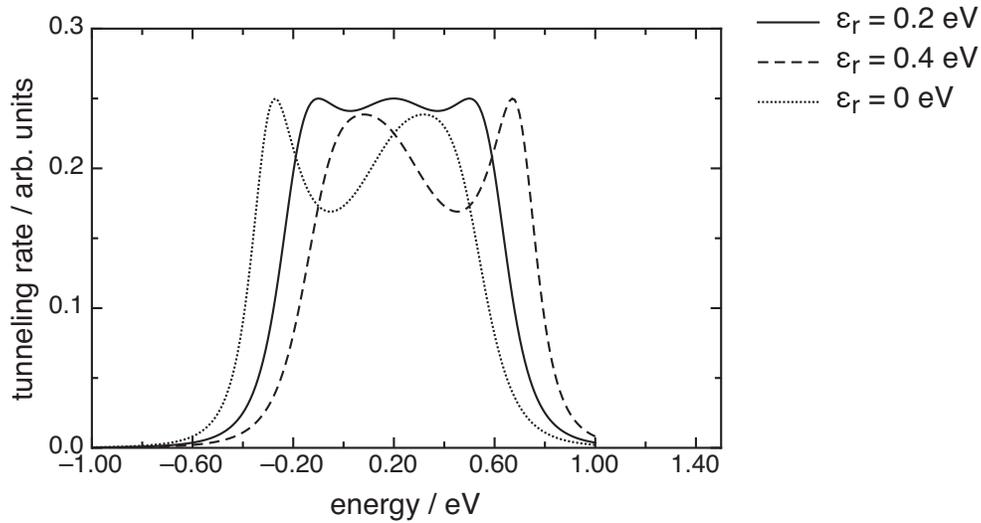}
\caption{Transition probability as a function of the energy $\epsilon$ of the tunneling electron for various values of $\epsilon_r$; the solvent coordinate $q$ was set to zero. System parameters: : $\epsilon_1 = 0.2$~eV, $v=0.1$~eV. }
\label{tstep}
\end{center}
\end{figure}

The transition probability $t(\epsilon, q)$ has local maxima, whenever $\epsilon$ is near one of the molecular orbitals. Since is depends only on the combination $\epsilon_r - 2 \lambda q$, it is sufficient to investigate the dependence on $\epsilon_r$ for $q=0$, as is done in Fig. \ref{tstep}. The interesting region lies where $\epsilon_r \approx \epsilon_i$. When both are equal, there are three distinct maxima
at the three molecular orbitals. When they are separated, only two maxima occur at the position of the atomic orbitals, since the splitting is too small to show up -- it is hidden beneath the maxima. 

\begin{figure}[htb]
\begin{center}
\includegraphics[height=7cm]{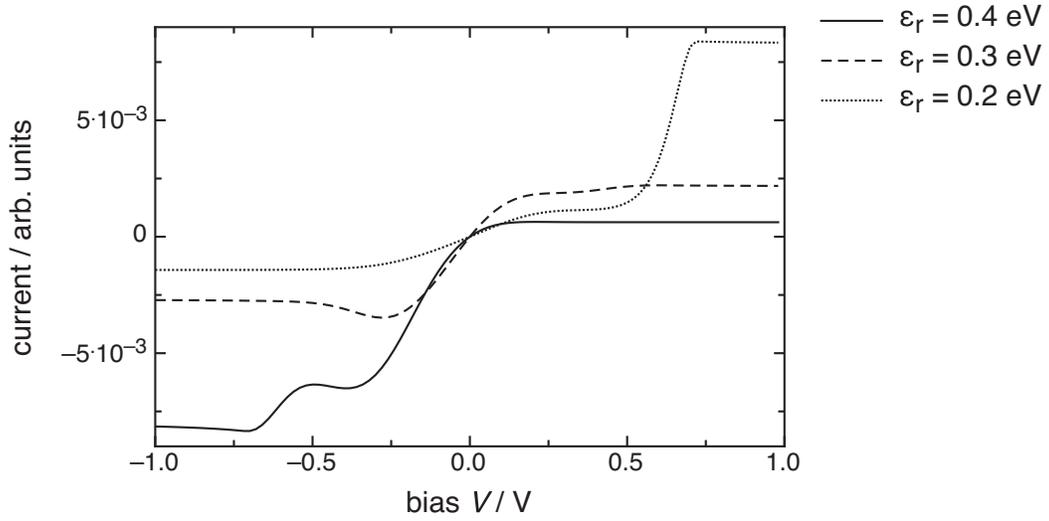}
\caption{Current-potential curves for various values of $\epsilon_r$. System parameters: : $\epsilon_1 = 0.2$~eV, $v=0.1$~eV. }
\label{steps}
\end{center}
\end{figure}

\begin{figure}[htb]
\begin{center}
\includegraphics[height=7cm]{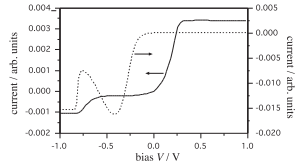}
\caption{Current-potential curves for small coupling to the leads. System parameters: $\Delta = 0.1$~eV,  $\epsilon_r = 0.5$~eV, $v=0.1$~eV, $\epsilon_1 = -0.2$~eV (left curve) and $\epsilon_1 = 0.3$~eV (right curve). }
\label{ndr}
\end{center}
\end{figure}

These oscillations in the transition probability  give rise to interesting current-potential curves exhibiting several steps and even regions with a negative differential resistance (see Fig. \ref{steps}), effects which do not occur with a single electronic intermediate state. The exact shape of these characteristics is determined by an interplay of three effects: The change of the potential-energy curves with the bias, the dynamic changes in the energy of the molecular orbitals as the solvent coordinate $q$ fluctuates, and the resulting  oscillations in the transition rate. These highly nonlinear effects are more pronounced when the coupling $\Delta$ to the two leads is weaker. Figure \ref{ndr} shows two examples where the system parameters have been chosen such that the curves either exhibit nice plateaus or a pronounced negative differential resistance.

\begin{figure}[htb]
\begin{center}
\includegraphics[height=7cm]{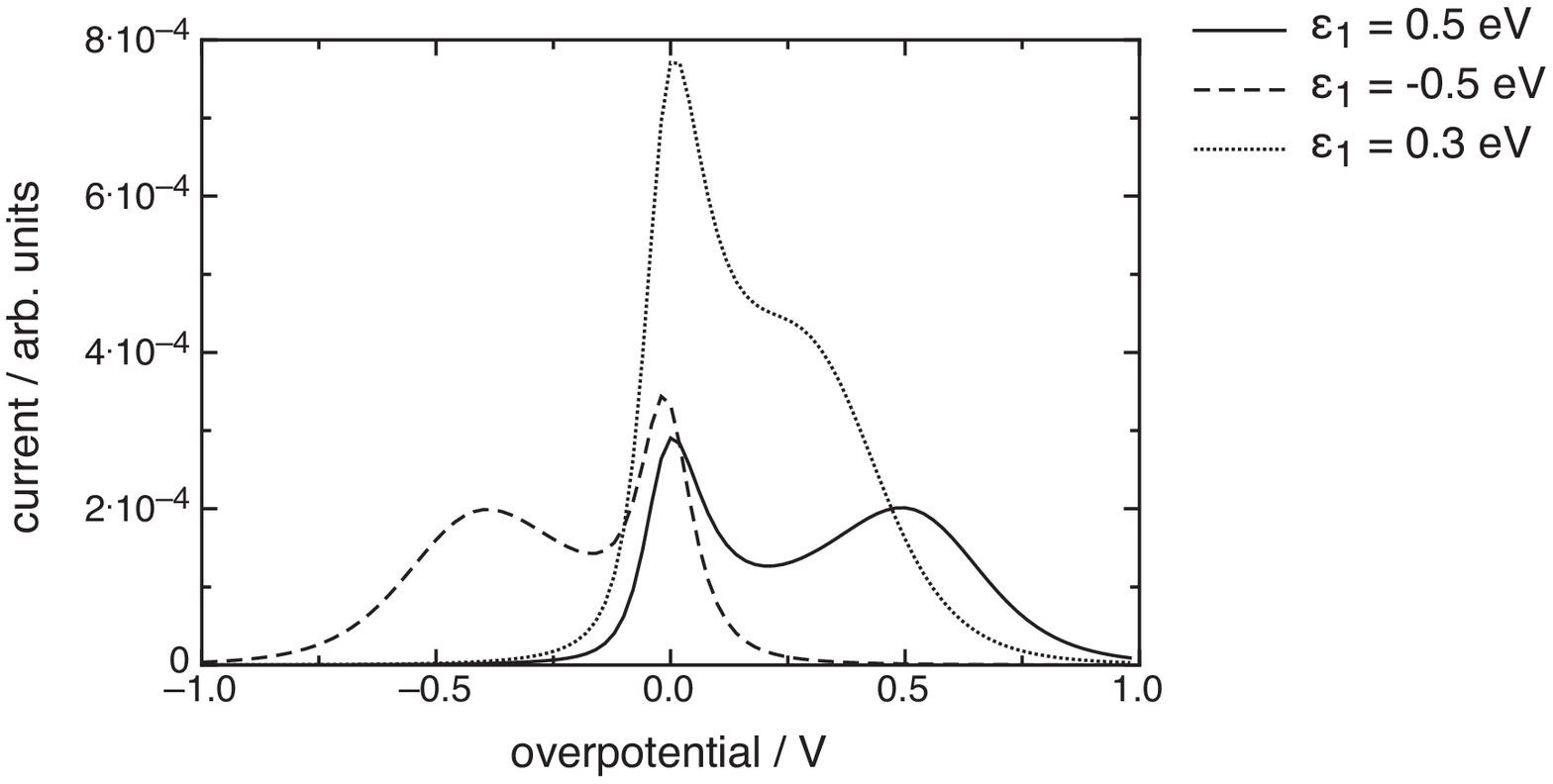}
\caption{Current at constant bias $V = 0.1$~V as a function of the overpotential $\eta$ for various values of $\epsilon_1^0$, the value of $\epsilon_1$ for vanishing overpotential;  $v=0.1$~eV. }
\label{eta}
\end{center}
\end{figure}

As pointed out in the introduction, in electrochemical systems two voltages can be controlled independently, the bias and the potential between the solution and one electrode. This makes it possible to perform spectroscopy of the electronic states in the wire, which experience the potential of the solution. We introduce the overpotential $\eta$ of the redox couple with respect to the right electrode through $\epsilon_r = \lambda - e_0 \eta$, and let the intermediate state shift in the same way: $\epsilon_1 = \epsilon_1^0 - e_0 \eta$. In a real system, because of the finite conductivity of the solution, $\eta$ may be only a fraction of the externally applied potential, but this would require only a trivial modification. If we keep the bias constant at a comparatively small value and scan the overpotential $\eta$, we obtain a peak in the current every time an electronic state lies within the tunneling range of energy between the two Fermi levels. A few examples are seen in Fig. \ref{eta}. The redox level always gives a peak near $\eta =0$, and for the parameters chosen we see a second peak near $\epsilon_1^0$. When these two energies lie close, one peak may appear as a shoulder. Note that the curves  for $\epsilon_1^0 = \pm 0.5$~eV in the figure are not quite symmetric, because the bias breaks the symmetry. In theory, we could expect to see up to three peaks in these curves corresponding to the three molecular orbitals formed, but these only occur for a very strong coupling $v$ and small energies of reorganisation. Otherwise the splitting induced by $v$ is hidden under the peak for $\epsilon_1$.

\section{Conclusions}
In this work we have presented a model for the conductivity of a molecular wire containing a redox system, and embedded in an electrochemical  environment. We considered the interaction of the redox system with a classical solvent, whose state was represented by a solvent coordinate $q$ in the spirit of the Marcus theory.  Using the wide-band approximation, we were able to derive an exact expression for the quantum conductance of a chain of arbitrary length. The thermal average over the solvent configurations had to be performed numerically.

Explicit calculations have been performed for a chain of three atoms. When the electronic levels of the neighboring atoms interact weakly with the redox couple -- because their energies are very different or the coupling is weak -- the wire behaves much like a single intervening atoms. Interesting new features arise when the redox couple interacts strongly with the neighboring levels. Since the redox level fluctuates with the solvent, this interaction is dynamic and changes with the solvent configuration. 
This gives rise to interesting relations between the current through the wire and the applied bias. In particular, rectification, extended current plateaus, and negative differential resistances can be observed. 

In an electrochemical environment two potential differences, the bias and the electrochemical potential of the wire, can be varied independently. This makes it possible to perform electronic spectroscopy at constant bias by changing the electrochemical potential. Intermediate states show up as characteristic current peaks.

Our treatment has been limited to a redox couple interacting with a classical solvent. An extension to the case where quantum modes couple to the electron transfer should be possible, using Green's function techniques that have been applied to the case of a single atom \cite{me2}. This could give rise to additional structure in current-potential curves.

\subsection*{Acknowledgement}
We gratefully acknowledge support by the DAAD (Germany) and DST (India).

\appendix
\section{Calculation of the determinants $d_{n}$}

 The method for obtaining the determinants  $d_{n}$ is discussed in this section. Let the determinant of a $n \times  n$ tri-diagonal matrix with diagonal entries $\epsilon$ and sub-diagonal entries $\upsilon$ be denoted as $d_{n}$. Then the following recursion holds good:
\begin{equation}
\label{a1}
M = \left( \begin{array}{ccc}
\epsilon & \upsilon & 0 \\
\upsilon  & \epsilon & \upsilon \vdots \\
\vdots    & \vdots   & \vdots \\
\vdots    & \upsilon \qquad  \epsilon \qquad \upsilon & \vdots \\
\vdots & \vdots & \vdots  \\
\vdots & \vdots & \upsilon \qquad \epsilon \\
\end{array} \right)_{n \times n} 
\end{equation}
\begin{equation}
\label{a2}
d_{n} = \epsilon d_{n-1} - \upsilon^{2} d_{n-2} 
\end{equation}

The above recursive relation can be written compactly in a matrix form:
\begin{equation}
\label{a3}
\left( \begin{array}{l}
d_{n} \\
d_{n-1} \\
\end{array} \right) =
\left( \begin{array}{ll}
\epsilon & -\upsilon^{2} \\
1 & 0 \\
\end{array} \right) \left( \begin{array}{l}
d_{n-1} \\
d_{n-2} \\
\end{array} \right)
\end{equation}
Denoting the recursion matrix as {\bf R}, it's eigenvalues   $\lambda_{1}$ and $\lambda_{2}$ and the matrix {\bf S} which diagonalises {\bf R} can be found easily. Iterating the recursion relation while noting that $d_{0} = 1$ and $d_{1} = \epsilon$, gives the following result:
\begin{equation}
\label{a4}
\left( \begin{array}{l}
d_{n} \\
d_{n-1} \\
\end{array} \right) = {\bf S}
\left( \begin{array}{ll}
\lambda_{1}  &   0  \\
0     &   \lambda_{2} \\
\end{array} \right)^{n-1}  {\bf S}^{-1} \left( \begin{array}{l}
d_{1} \\
d_{0}  \\
\end{array} \right) 
\end{equation}

 The final required expression for $d_{n}$ obtained from the above equation  is shown below:
 \begin{equation}\label{a5}
d_{n}  = \frac{\lambda_{1} ^{n+1} - \lambda_{2} ^{n+1} }{\lambda_{1} - \lambda_{2}} \quad \mbox{with} \quad 
\lambda_{1,2}  = \frac{ \epsilon \pm \sqrt{ \epsilon - 4\upsilon^{2}}}{2} 
\end{equation}

The methods discussed above can be use d to get the inverse elements of the matrix:
\begin{equation}
\label{a6}
 M^{-1} _{11} = \frac{d_{n-1}}{d_{n}}    \quad M^{-1} _{12} = \frac{-\upsilon d_{n-2}}{d_{n}} \quad
M^{-1} _{1j} = \frac{ (-\upsilon)^{j-1} d_{n-j}}{d_{n}} 
\end{equation}

\newpage
\subsection*{Figures}

\newpage



\newpage

\begin{thebibliography}{1}
\bibitem{Jort_ed} J. Jortner, M. Ratner, Eds. {\it Molecular Electronics} Blackwell Science, Cambridge, MA, (1997)
\bibitem{Aviram_ed} A. Aviram, M. Ratner, Eds. {\it Molecular Electronics: Science and Technology} The New york Academy of Sciences, New York (1998) Vol. 2. 
\bibitem{Adams01} D.M. Adams . {\it et al}, J. Phys. Chem B. {\bf 107}, 6668-6697 (2003).
\bibitem{lent01} C.S. Lent and P.D. Tougaw,   Proc IEEE {\bf 97}  541-557 (1997).
\bibitem{Amlani01} I. Amlani {\it et . al},  Science {\bf 284} 289-91 (1999).
\bibitem{Kum01} R kummamuru {\it et . al},   App. Phys. Lett {\bf 81} 1332-4 (2002).
\bibitem{Lieberman01} M. Lieberman {\it et. al},  Ann. N. Y. Acad. Sci {\bf 960} 225-29 (2002).
\bibitem{cygan01} M. T. Cygan {\it et . al},  J. Am. Chem. Soc {\bf 103} 8122-7 (1998).
\bibitem{Decher01} G. Decher,  Science {\bf 277} 1232-7 (1997) .
\bibitem{Newton01} M.D. Newton, Chem. Rev. {\bf 91} 767-92. (1991).
\bibitem{Davis01} W. B. Davis {\it et. al},  J. Phys. Chem . A {\bf 101} 6158-64 (1997).
\bibitem{Marcus01} P. Siddharth and R.A. MArcus, J. Phys. Chem {\bf 90} 2985-9 (1990).
\bibitem{McConnell01} H. M. McConnell, J. Chem. Phys {\bf 35}  508 (1961).
\bibitem{Mujica01} V. Mujica, M. Kemp and M. A. Ratner, J. Chem. Phys {\bf 101} 6849 (1994).
\bibitem{Mujica02} V. Mujica, M. Kemp and M. A. Ratner, J. Chem. Phys {\bf 101} 6856 (1994).
\bibitem{Mujica03} V. Mujica, A. E. Roitberg and M. A. Ratner, J. Chem. Phys {\bf 112} 6834-9 (2000). 
\bibitem{caroli} C. Caroli, R. Combescot, P. Nozieres and D. Saint-James, J. Phys. C. {\bf 4}, 916 (1971).
\bibitem{wingreen1} N. S. Wingreen, K. W. Jacobsen and J. W. Wilkins, Phys. Rev. Lett. {\bf 61} 1396-1399 (1988).
\bibitem{meir1} Y. S.  Meir ,and N. S. Wingreen,Phys. Rev. Lett. {\bf 68},  2512-15 (1992).
\bibitem{meir2} A. -P. Jauho, N. S.  Wingreen, and Y. S. Meir,   Phys. Rev. B. {\bf  50}  5528-44 (1994).

\bibitem{Hu01} G.Y. Hu and R. F. O'Connell, Phys. Rev. B {\bf 36} 5798 (1987).
\bibitem{Camalet01} S. Camalet {\it et. al} Phys. Rev. lett. {\bf 90} 210602 (2003).
\bibitem{Camalet02} S. Camalet {\it et. al} Phys. Rev. B. {\bf 70} 155326 (2004).
\bibitem{Segal01} D. Segal, A. Nitzan and P. H\"{a}nggi, J. Chem. Phys. {\bf 119} 6840 (2003).
\bibitem{sen01} D. Sen and A. Dhar, Phys. Rev. B {\bf 73} 085119-51 (2006).
\bibitem{Brand01} M. Brandbyge {\it et. al}, Phys. Rev. B {\bf 65} 165401 (2002).
\bibitem{Taylor01} J. Taylor {\it et al}, Phys. Rev. B {\bf 63} 245407 (2001).
\bibitem{Taylor02} J. Taylor {\it et. al} Phys. Rev. Lett {\bf 89} 138301 (2002).
\bibitem{DiVentra01} M. DiVentra, S. T. Pantelides and N. D. Lang, Phys. Rev. Lett {\bf 84} 979 (2000).
\bibitem{Datta02} P.S. Damle, A. W. Ghosh and S. Datta Phys. Rev. B {\bf 64} 201403 (2001). 
\bibitem{Lundin01} U. Lundin and H. McKenzie, Phys. Rev. B {\bf 66} 075303 (2002).
\bibitem{Zhu01} J. X. Zhu and A. V. Balatsky, Phys. Rev. B {\bf 67} 165326 (2003).
\bibitem{Alexandrov01} A. S. Alexandrov, A. M. Bratkovsky and R. S. Williams, Phys. Rev. B {\bf 67} 075301 (2003).
\bibitem{Alexandrov02} A. S. Alexandrov and A. M. Bratkovsky, Phys. Rev. B {\bf 67} 235312  (2003).
\bibitem{Flensberg01} K. Flensberg, Phys. Rev. B {\bf 68} 205323 (2003).
\bibitem{Mozyrsky01} D. Mozyrsky, M. B. Hastings and I. Martin, Phys. Rev. B {\bf 73} 035104 (2006).


\bibitem{me1}  W. Schmickler,  J. Electroanal.\ Chem.\ {\bf 296} (1990) 283. 
\bibitem{cindra}  W. Schmickler and C. Widrig, 
 J. Electroanal.\ Chem.\ {\bf 336} (1992) 213.
 \bibitem{ulstrup1}A.M. Kuznetsov, P. Sommer-Larsen, and J. Ulstrup, Surf. Sci. \textbf{275} (1992) 52.
\bibitem{me2}  W. Schmickler, Surf.\ Science {\bf
  295} (1993) 43 


\bibitem{Tao01} N. J. Tao, Phys. Rev. Lett. {\bf 76} 4066 (1996)
\bibitem{Han01} W. H. Han, E. N. Durantini, T. A. Moore, D. Gust, P. Rez, G. Leatherman, G. R. Seely, N. Tao and S. M. Lindsay, J. Phys. Chem. B. {\bf 101} 10719 (1997).
\bibitem{Holmin01} R. E. Holmin, R. F. Ismagilov, R. Haag, V. Mujica, M. A. Ratner, M. A. Ramp and G. M. Whitesides, Angew. Chem. Int. Edn {\bf 40} 2316 (2001).
\bibitem{Tran01} E. Tran, M. A. Rampi and G. M. Whitesides, Angew. Chem. Int. Edn {\bf 43} 3835(2004).
\bibitem{Xiao01} X. Y. Xiao, L.A. Nagahara, A. M. Rawlett and N. Tao,  J. Am. Chem. Soc {\bf 127} 9235 (2005).
\bibitem{He01} J. He and S. M. Lindsay J. Am. Chem. Soc {\bf 127} 11932 (2005).
\bibitem{Albrecht01} T. Albrecht, K Moth-Poulsen, J. B. Christensen, J. Hjelm, T. Bjornholm and J. Ulstrup,  J. Am. Chem. Soc. {\bf 128} 6574 (2006).
\bibitem{Chi01} Q. Chi, J Zhang, P. S. Jensen, H. E. M. Christensen and J. Ulstrup, Faraday Dicuss. {\bf 131} 181 (2006).
\bibitem{Albrecht02} T. Albrecht , K. Moth-Poulsen, J. B. Christensen, A. Guckian, T. Bjornholm, J. G. Vos and J. Ulstrup, Faraday Dicuss. {\bf 131} 265 (2006).
\bibitem{Li01} Z. Li, H. Han, G. Mesazaros, I. Pobelov, Th. Wandlowski, A. Blaszczyk and M. Mayor, Faraday Discuss. {\bf 131} 121 (2006).
\bibitem{Alessandrini01} A. Alessandrini, S. Corni and P. Facci, Phys. Chem. Chem. Phys. {\bf 8} 4383 (2006). 
\bibitem{galperin} M. Galperin, A. Nitzan, and M.A. Ratner, J. Phys. Condens. Matter \textbf{20} (2008) 374107.
\bibitem{medved}A. M. Kuznetsov and I. G. Medvedev, J. Phys. Condens. Matter \textbf{20} (2008) 374112..





\bibitem{met1}  W. Schmickler, Electrochim.\ Acta, \textbf{41} 
 (1996) 2329.




\bibitem{Datta1} W. Tian, S. Datta, S. Hong, R. Reifenberger, J. I. Henderson, and C. I. Kubiak, J. Chem. Phys. {\bf 109}, 2874 (1998).
\bibitem{evenson} J. Evenson and M. Karplus, J. Chem. Phys. {\bf 96}, 5272, (1992).

\bibitem{sashame}
A.N. Kuznetsov and W. Schmickler, Chem. Phys. \textbf{282} (2002) 371.
\end{thebibliography}
\end{document}